\documentclass[journal]{IEEEtran}
\usepackage[noadjust]{cite}
\usepackage[caption=false,font=footnotesize]{subfig}
\usepackage{stfloats}
\usepackage{url}
\usepackage[colorlinks=true,linkcolor=black,citecolor=blue,urlcolor=blue]{hyperref}
\usepackage{amssymb,amsmath,amsthm}
\usepackage{array}
\usepackage{amsfonts}
\usepackage{graphicx}
  \graphicspath{{./pics/}}
  \DeclareGraphicsExtensions{.pdf,.png}
\usepackage{xspace}
\usepackage{pgf}
\usepackage{tikz}
  \usetikzlibrary{arrows,shapes,fit,positioning}
  \usetikzlibrary{automata,backgrounds}
  \usetikzlibrary{decorations.pathmorphing}
  \usetikzlibrary{calc}
  \usetikzlibrary{positioning}
  \usetikzlibrary{decorations.markings}
  \usetikzlibrary{intersections}
\usepackage[T1]{fontenc}
\usepackage[latin1]{inputenc}
\usepackage[english]{babel}
\usepackage{booktabs}
  \newcommand{\ra}[1]{\renewcommand{\arraystretch}{#1}}
\usepackage{graphbox}
\usepackage{outlines}
\usepackage{algpseudocode}
\usepackage[linesnumbered,ruled,noend,noline]{algorithm2e}
  
\usepackage{fixmath}
\usepackage[export]{adjustbox}
\usepackage[bottom]{footmisc}
\usepackage{empheq,fancybox}
\usepackage[textwidth=1.9cm,color=green!10,textsize=footnotesize]{todonotes}

\allowdisplaybreaks

\newtheorem{thm}{Theorem}
\newtheorem{cor}[thm]{Corollary}
\newtheorem{lem}[thm]{Lemma}
\newtheorem{remark}[thm]{Remark}
\newtheorem{defn}[thm]{Definition}
\newtheorem{exmp}[thm]{Example}

\DeclareMathOperator{\supC}{\rm supC}
\DeclareMathOperator{\supN}{\rm supN}
\DeclareMathOperator{\supCN}{\rm supCN}
\newcommand{\eps}{\varepsilon}
\newcommand{\complclass}[1]{{\scshape #1}\xspace}
\newcommand{\PSpace}{\complclass{PSpace}}
\newcommand{\NP}{\complclass{NP}}

\begin{document}

\title{Supervisory Control of Modular Discrete-Event Systems under Partial Observation: Normality}%

\author{Jan Komenda and Tom{\' a}{\v s}~Masopust
  \thanks{J. Komenda and T. Masopust are with the Institute of Mathematics of the Czech Academy of Sciences, Prague, Czechia, and with the Faculty of Science, Palacky University Olomouc, Czechia.
  Emails: {\tt komenda@ipm.cz}, {\tt tomas.masopust@upol.cz}.}%
}

\markboth{}{}

\maketitle

\begin{abstract}
  Complex systems are often composed of many small communicating components called modules. We investigate the synthesis of supervisory controllers for modular systems under partial observation that, as the closed-loop system, realize the supremal normal sublanguage of the specification. We call such controllers maximally permissive normal supervisors.
  The challenge in modular systems is to find conditions under which the global nonblocking and maximally permissive normal supervisor can be achieved locally as the parallel composition of local normal supervisors. We show that a structural concept of hierarchical supervisory control called modified observation consistency (MOC) is such a condition.
  However, the algorithmic verification of MOC is an open problem, and therefore it is necessary to find easily-verifiable conditions that ensure MOC. We show that the condition that all shared events are observable is such a condition.
  Considering specifications, we examine both local specifications, where each module has its own specification, and global specifications.
  We combine our results for normality with the existing results for controllability to locally synthesize the nonblocking and maximally permissive controllable and normal supervisor.
  Finally, we illustrate the results on an industrial case study of the patient table of an MRI scanner.
\end{abstract}

\section{Introduction}
  Organizing large-scale complex systems into interconnected communicating components is a common engineering practice, which has many applications across control theory and computer science, including manufacturing, robotics, and artificial intelligence~\cite{BaierM15,SyllaLRD18,Raisch2005}.

  The key concepts of supervisory control under partial observation are controllability and observability~\cite{CL08,Won04}. Unlike controllability, however, observability is not closed under union, and hence the set of observable sublanguages of a given language does not have the supremal element. Therefore, other concepts, including normality or relative observability, are used instead of observability~\cite{CaiZW15,AlvesCB17}. For more results on supervisory control of partially-observed (monolithic) plants, we refer the reader to, e.\,g., Takai and Ushio~\cite{TakaiUshio2003}, Thistle and Lamouchi~\cite{Thistle09}, or Yin and Lafortune~\cite{YinLafortuneTAC16}.

  Compared with the monolithic plant, modular discrete-event systems (modular DES) consist of a concurrent composition of many small components, also known as modules. The main challenge of supervisory control for modular DES is the problem how to synthesize local supervisors in such a way that the concurrent behavior of local modules controlled by the corresponding local supervisors coincides with the behavior of the global (monolithic) plant controlled by the nonblocking and maximally permissive supervisor.
  The idea of a local controller synthesis is natural and has been discussed in the literature since the early days of supervisory control~\cite{RW88}.

  In this paper, we use the concept of normality. Normality is stronger than observability, and if all controllable events are observable, then the two notions coincide~\cite{CL08,Won04}.

  Given a modular DES and a specification, the problem now is how to compute the supremal controllable and normal sublanguage of the specification without explicitly constructing the global plant. Since local computations of the supremal controllable sublanguage have widely been investigated in the literature, we focus on the local construction of the supremal normal sublanguage; and we show how to combine the results.

  Although the supervisor synthesis for a monolithic plant with complete observation is polynomial in time, the main motivation for the computation of local supervisors is the avoidance of the construction of the monolithic plant, the size of which grows exponentially in the number of local modules. Furthermore, unlike complete observation, there are no polynomial-time algorithms for the synthesis of supervisors for systems under partial observation, and hence the local computation of supervisors for systems under partial observation is even more important.

\subsection*{Systems with Complete Observation}
  We now briefly review the main approaches to supervisory control of modular DES under complete observation. In this case, the problem is to locally compute the supremal controllable sublanguage of a global specification without explicitly constructing the global plant.

  De Queiroz and Cury~\cite{Queiroz2000} have considered modular DES, where local modules have no events in common, equipped with a set of specifications. The specifications do not correspond to the modules, and therefore, for each specification, a local plant is constructed as a parallel composition of those modules that share an event with the specification. De Queiroz and Cury have shown that the parallel behavior of the maximally permissive local supervisors, computed for each specification and the corresponding local plant, coincides with the behavior of the nonblocking and maximally permissive monolithic supervisor.

  Considering the same framework, a similar approach was discussed by Hill and Tilbury~\cite{HillTilbury2006}, who further employed an abstraction and a structuring into several levels.

  Gaudin and Marchand~\cite{Gaudin2007}, on the other hand, have considered a global prefix-closed specification, which they localize with the help of inverse projections to the alphabets of local modules. Then, they employ the concept of partial controllability to compute maximally permissive local supervisors. The parallel composition of the constructed local supervisors is then maximally permissive in the monolithic sense under the condition that all shared events are controllable.

  Willner and Heymann~\cite{WILLNER1991} have investigated global decomposable prefix-closed specifications---specifications that can be decomposed according to the alphabets of local modules. They have shown that if all shared events are controllable, then the parallel composition of maximally permissive local supervisors is maximally permissive in the monolithic sense.

  However, the assumption of decomposability of the specification~\cite{WILLNER1991} or of the solution~\cite{JiangKumar2000,Rohloff2006} is very restrictive.
  Therefore, we introduced the notion of {\em conditional decomposability}~\cite{KMvS2011}; intuitively, rather than to decompose the specification with respect to the alphabets of local modules, we search for a coordinator that covers the shared communication among the modules in such a way that the specification can be decomposed with respect to the alphabets of local modules composed with the coordinator. In this way, we reduce a modular DES with a global specification to a modular DES with local specifications~\cite{JDEDS15}; see also Section~\ref{subsectionCMDES}.

  Flordal et al.~\cite{FlordalFabian07} have developed an incremental compositional technique that avoids the construction of the monolithic supervisor and preserves the nonblockingness and maximal permissiveness of the closed-loop system.
  We further refer the reader to Abdelwahed and Wonham~\cite{AbdelwahedWonham2002} and Lee and Wong~\cite{Lee2002}.

  Another approach to supervisory control of modular systems is based on the {\em observer property}~\cite{WW96}, on the property of {\em output control consistency\/} (OCC)~\cite{ZhongW1990}, and on the property of {\em local control consistency\/} (LCC)~\cite{SB11}; these properties are concepts of hierarchical supervisory control. Namely, given a modular DES with a decomposable prefix-closed specification, if each projection from the overall alphabet to the alphabet of a local module satisfies the observer property and the LCC (or OCC) condition, then the parallel composition of maximally permissive local supervisors coincides with the monolithic nonblocking and maximally permissive supervisor, see Section~\ref{consequences} for more details.
  Feng~\cite{FLT} has lifted this result to non-prefix-closed specifications by the assumption that the local supervisors are nonconflicting.

  Combining this result with the result of Willner and Heymann~\cite{WILLNER1991} then gives that, for a modular DES under complete observation with a decomposable specification, if all shared events are controllable, then the concurrent behavior of maximally permissive local closed-loop systems coincides with the behavior of the nonblocking and maximally permissive monolithic closed-loop system.

\subsection*{Systems with Partial Observation}
  We investigate the synthesis of supervisory controllers for modular DES under partial observation. In particular, we focus on the synthesis of maximally permissive {\em normal supervisors}, which are supervisors realizing the supremal normal sublanguage of the specification.

  Whereas the synthesis of maximally permissive local supervisors is well understood for modular DES under complete observation, the situation for partial observation is significantly different, and only a few partial results can be found in the literature.

  Komenda and Van Schuppen~\cite{KvS_TCS07} have shown that the computation of local normal supervisors is maximally permissive for prefix-closed specifications if the modules share no events. They have further suggested a condition of mutual normality, which is, however, too restrictive~\cite{KvS_IEEE08}.
  Komenda et al.~\cite{scl2011} have extended the coordination control framework to partial observation, where additional conditions are required to achieve maximal permissiveness of the composition of local supervisors.

  Among other works related to supervisory control of modular DES, Rohloff and Lafortune~\cite{Rohloff2006}, Su and Lennartson~\cite{SL2017}, or Liu et al.~\cite{LKML22} have modelled multi-agent systems as a modular DES composed of isomorphic modules. The isomorphism of modules allows us to handle the number of agents that can dynamically decrease or increase without even an upper bound.

  In this paper, we show that a concept of hierarchical supervisory control called {\em modified observation consistency\/} (MOC) is of interest for modular DES; namely, if a modular system satisfies the MOC condition, then the global nonblocking and maximally permissive normal supervisor can be achieved locally as the parallel composition of nonblocking and maximally permissive local normal supervisors (Theorem~\ref{thm15}).

  However, the algorithmic verification of MOC is a challenging open problem; the problem is known to be \PSpace-hard, but unknown to be decidable~\cite{firstPart}. Therefore, we need other, easily-verifiable conditions that ensure MOC. We show that the assumption that {\em all shared events are observable\/} is such a condition (Theorem~\ref{cor_conjecture_proven}).

  The reader may notice that several of the above-discussed approaches to supervisory control of modular DES under complete observation make a similar assumption; namely,  that {\em all shared events are controllable}. Consequently, our condition that all shared events are observable naturally extends and complements the existing results.

  Concerning the structure of a specification, we discuss two fundamental cases: (i) the specification is given as a set of local specifications, where each local module has its own local specification, and (ii) the specification is global, given as a sublanguage of the global plant language. For the latter case, we employ our coordination control framework~\cite{JDEDS15} and show that under the assumption that {\em all coordinated events are observable}, global and local computations of nonblocking and maximally permissive normal supervisors coincide (Theorem~\ref{mainCC}).

  Last but not least, our results provide further evidence that the modular supervisory control framework benefits from the results and concepts of hierarchical supervisory control. In particular, the concepts of the observer property, OCC, LCC, and MOC conditions, developed for supervisory control of hierarchical systems, find applications in supervisory control of modular DES.

\section{Preliminaries and Definitions}
  We assume that the reader is familiar with the basic concepts of supervisory control~\cite{CL08}. For a set $A$, we denote by $|A|$ the cardinality of $A$. For an alphabet (finite nonempty set) $\Sigma$, we denote by $\Sigma^*$ the set of all finite strings over $\Sigma$; the empty string is denoted by $\eps$. A language $L$ over $\Sigma$ is a subset of $\Sigma^*$. The prefix closure of a language $L$ is the set $\overline{L}=\{w\in \Sigma^* \mid \text{there exists } v\in\Sigma^* \text{ such that } wv\in L\}$. A language $L$ is prefix-closed if $L=\overline{L}$.

  A {\em projection\/} $R\colon \Sigma^* \to \Gamma^*$, where $\Gamma \subseteq \Sigma$ are alphabets, is a morphism for concatenation defined by $R(a) = \eps$ if $a\in \Sigma-\Gamma$, and $R(a) = a$ if $a\in \Gamma$. The action of $R$ on a string $a_1a_2\cdots a_n$ is to remove events that are not in $\Gamma$, that is, $R(a_1a_2\cdots a_n)=R(a_1)R(a_2)\cdots R(a_n)$. The inverse image of a string $w\in\Gamma^*$ under the projection $R$ is the set $R^{-1}(w)=\{s\in \Sigma^* \mid R(s) = w\}$. The definitions can readily be extended to languages.

  A {\em deterministic finite automaton\/} (DFA) is a quintuple $G = (Q,\Sigma,\delta,q_0,F)$, where $Q$ is a finite set of states, $\Sigma$ is an alphabet, $q_0 \in Q$ is the initial state, $F\subseteq Q$ is the set of marked states, and $\delta \colon Q\times\Sigma \to Q$ is the transition function that can be extended to the domain $Q\times \Sigma^*$ in the usual way. 
  The language $L(G) = \{w\in \Sigma^* \mid \delta(q_0,w)\in Q \}$ is {\em generated\/} by $G$, while the language $L_m(G) = \{w\in \Sigma^* \mid \delta(q_0,w) \in F\}$ is {\em marked\/} or {\em accepted\/} by $G$.
  By definition, $L_m(G)\subseteq L(G)$, and $L(G)$ is prefix-closed. If $\overline{L_m(G)} = L(G)$, then $G$ is called {\em nonblocking}.

  A discrete-event system (DES) over $\Sigma$ is a DFA over $\Sigma$ together with the determination of {\em controllable events} $\Sigma_c\subseteq \Sigma$ and {\em uncontrollable events} $\Sigma_{uc} = \Sigma - \Sigma_c$, and of {\em observable events} $\Sigma_o\subseteq \Sigma$ and {\em unobservable events} $\Sigma_{uo} = \Sigma - \Sigma_o$.

  For a DES $G$ over $\Sigma$, we denote the set of control patterns by $\Gamma = \{\gamma \subseteq \Sigma \mid \Sigma_{uc} \subseteq \gamma\}$, and the projection removing unobservable events by $P\colon \Sigma^* \to \Sigma_o^*$.
  A {\em supervisor\/} of $G$ with respect to a set of control patterns $\Gamma$ is a map $S\colon P(L(G)) \to \Gamma$ returning, for every observed string, a set of enabled events that always includes all uncontrollable events.
  The {\em closed-loop system\/} of $G$ under the supervision of $S$ is the minimal language $L(S/G)$ such that $\eps \in L(S/G)$ and for every $s\in L(S/G)$, if $sa \in L(G)$ and $a \in S(P(s))$, then $sa \in L(S/G)$. If the closed-loop system is nonblocking, that is, $\overline{L_m(S/G)}=L(S/G)$, the supervisor $S$ is called {\em nonblocking}. Intuitively, the supervisor disables some of the controllable transitions based on the partial observation of the system.

  There are two views on the marked language of the closed-loop system: (i) the marking is adopted from the plant $G$, that is, $L_m(S/G) = L(S/G) \cap L_m(G)$, and (ii) the supervisor marks according to a given specification $M\subseteq L(G)$, that is, $L_m(S/G)=L(S/G)\cap M$. 
  In the latter case, the existence of a supervisor achieving the specification is equivalent to controllability and observability of the specification, whereas, in the former case, an additional assumption of $L_m(G)$\mbox{-}closedness is needed~\cite[Section~6.3]{Won04}. 

  However, unlike controllability, observability is not closed under union, and hence the set of observable sublanguages of a given language does not have the supremal element. For this reason, other concepts are used instead of observability. In this paper, we use the concept of normality. Normality is stronger than observability, and coincides with observability if all controllable events are observable~\cite{CL08,Won04}.

  For a DES $G$ over $\Sigma$, a language $K\subseteq L_m(G)$ is {\em controllable} with respect to the language $L(G)$ and the set $\Sigma_{uc}$ of uncontrollable events if $$\overline{K}\Sigma_{uc}\cap L(G)\subseteq \overline{K}\,.$$
  The language $K$ is {\em normal\/} with respect to the language $L(G)$ and the projection $P\colon\Sigma^*\to\Sigma_o^*$ if $$\overline{K} = P^{-1}[P(\overline{K})]\cap L(G)\,.$$

  For a prefix-closed language $L$ and a (not necessarily prefix-closed) specification $K\subseteq L$, we denote by $\supN(K,L,P)$ the supremal normal sublanguage of the specification $K$ with respect to the language $L$ and projection $P$~\cite{LinWon88}.
  
  The parallel composition of languages $L_i\subseteq \Sigma_i^*$ is the language $\|_{i=1}^{n} L_i = \cap_{i=1}^{n} P_i^{-1}(L_i)$, where $P_i \colon (\cup_{i=1}^{n} \Sigma_i)^* \to \Sigma_i^*$ is the projection, $i=1,\ldots,n$.
  A definition of the parallel composition for automata can be found in the literature~\cite{CL08}. In particular, for DFAs $G_i$, we have $L(\|_{i=1}^{n} G_i) = \|_{i=1}^{n} L(G_i)$ and $L_m(\|_{i=1}^{n} G_i) = \|_{i=1}^{n} L_m(G_i)$.
  The languages $L_i$ are {\em (synchronously) nonconflicting\/} if $\overline{\|_{i=1}^{n} L_{i}} = \|_{i=1}^{n} \overline{L_{i}}$.

  A modular DES $G = \|_{i=1}^n G_i$ consists of the parallel composition of $n\ge 2$ systems or modules $G_i$ over local alphabets $\Sigma_i$, for $i=1,\ldots,n$.

\subsection{Modular Supervisory Control}
  We now briefly review the principles and concepts of supervisory control of modular DES.

  The modular supervisory control problem consists of a modular system modeled by a set of $n\ge 2$ automata $$G_1,\ldots, G_n$$ generating languages $L_1=L(G_1),\ldots,L_n=L(G_n)$, respectively, with the global (monolithic) behavior $$L = \|_{i=1}^{n} L_i\,,$$ and of a specification $K$ given either
  \begin{enumerate}
    \item as a parallel composition $K=\|_{i=1}^{n} K_i$ of a set of local specifications $K_i \subseteq L_i$, or
    \item as a global specification $K \subseteq \|_{i=1}^{n} L_i$.
  \end{enumerate}

  The aim is to synthesize local controllers $S_i$ such that
  \begin{align}\label{eq1}
    \|_{i=1}^{n} L_m(S_i/G_i) = L_m(S/\|_{i=1}^{n} G_i)
  \end{align}
  where $S$ denotes the nonblocking and maximally permissive supervisor for the global specification $K$ and the global plant language $L$.

  In particular, the fundamental question is under which conditions \eqref{eq1} holds.

  This problem is well understood for modular systems under complete observation, where we essentially have two types of sufficient conditions ensuring the maximal permissiveness of the local (modular) control synthesis:
  \begin{enumerate}
    \item conditions adopted from hierarchical supervisory control; namely, the observer property and OCC/LCC conditions~\cite{feng08,SB11}, and
    \item the condition of mutual controllability and the variants thereof~\cite{KvS_TCS07}.
  \end{enumerate}

  For partially observed DES, however, only the second type of conditions was discussed in the literature~\cite{CDC19}.

  In this paper, we discuss a condition of the first type called modified observation consistency.

\subsection{Modified Observation Consistency}
  Modified observation consistency (MOC) is a concept of hierarchical supervisory control under partial observation developed to ensure hierarchical consistency~\cite{firstPart}.

\begin{figure}
    \centering
    \includegraphics[scale=1]{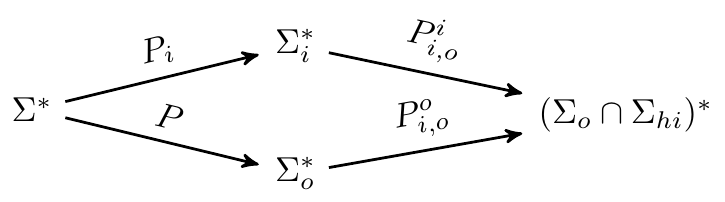}
    \caption{Commutative diagram of abstractions and projections.}
    \label{projections}
  \end{figure}

  Before we recall the definition of MOC, we fix the notation for projections. Namely, we denote system's partial observation by the projection $$P\colon \Sigma^* \to \Sigma^*_o\,,$$ the local projection to a module by the projection $$P_i\colon \Sigma^* \to \Sigma_{i}^*\,,$$ and the corresponding restricted observations and projections by
  \begin{center}
    $P^i_{i,o}\colon \Sigma_{i}^{*} \to (\Sigma_{i}\cap \Sigma_o)^*$ ~and~ $P^{o}_{i,o}\colon \Sigma_o^* \to (\Sigma_{i}\cap \Sigma_o)^*$,
  \end{center}
  see Figure~\ref{projections}.

  \begin{defn}\label{defMOC}
    A prefix-closed language $L \subseteq \Sigma^*$ is {\em modified observation consistent} (MOC) with respect to projections $P_i$, $P$, and $P^i_{i,o}$ if for every string $s\in L$ and every string $t' \in P_i(L)$ such that $P^i_{i,o}(P_i(s)) = P^i_{i,o}(t')$, there exists a string $s' \in L$ such that $P(s) = P(s')$ and $P_i(s') = t'$.
  \end{defn}
  
\begin{figure}
    \centering
    \includegraphics[scale=.9]{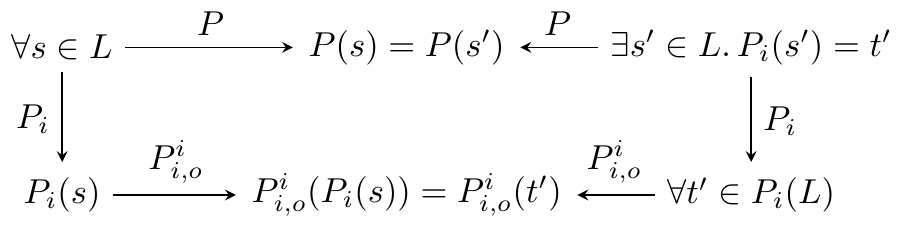}
    \caption{Illustration of the MOC condition.}
    \label{figMOC}
  \end{figure}

  Intuitively, every string of a localized plant that locally looks the same as a global string $s$ has a locally equivalent global string $s'$ that looks the same as $s$, cf.~Figure~\ref{figMOC} for an illustration.

  We define the set of local observable events as $\Sigma_i \cap \Sigma_o$.
  For simplicity, in the sequel, we denote the intersection of two alphabets by the corresponding subscripts separated by commas; for instance,
  \[
    \Sigma_{i,o}=\Sigma_i \cap \Sigma_o\,.
  \]

\begin{figure}
    \centering
    \includegraphics[scale=1]{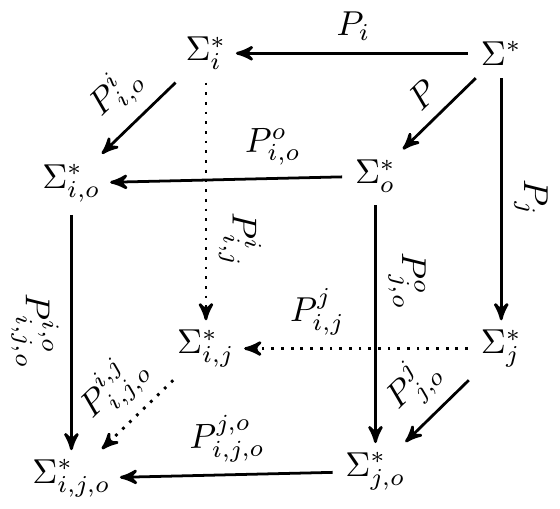}
    \caption{Our notation for the projections used in modular systems.}
    \label{fig1}
  \end{figure}

  We use the same notation for the intersection of other alphabets, and for the intersection of more than two alphabets. The corresponding projections that we use in this paper are summarized in Figure~\ref{fig1}.

  Finally, we define the set of all shared events of the modular DES $G=\|_{i=1}^{n} G_i$, where the alphabet of $G_i$ is $\Sigma_i$, as
  \begin{align}\label{sharedEvents}
    \Sigma_s = \bigcup_{i\neq j} (\Sigma_i\cap \Sigma_j)
  \end{align}
  and we assume that
  \begin{empheq}[box=\shadowbox]{gather*}
    \text{modular components agree on the observability status}\\
    \text{of shared events.}
  \end{empheq}

  In other words, if an event is observable in one component, it is observable in all components where it appears; formally,
  \[
    \Sigma_{i,o} \cap \Sigma_{j} = \Sigma_{i}  \cap  \Sigma_{j,o}= \Sigma_{i,o}  \cap  \Sigma_{j,o}\,.
  \]

\section{Synthesis of Normal Supervisors for Modular DES with Local Specifications}
\label{subsectionMDES}
  In this section, we consider the case of local specifications, where each module $L_i$ has its own specification $K_i$.
  The problem is to find conditions under which the parallel composition of the supremal normal sublanguages computed locally for each pair $(K_i,L_i)$ of a specification and a local plant coincides with the supremal normal sublanguage computed for the global specification $K = \|_{i=1}^{n} K_i$ and the monolithic plant $L = \|_{i=1}^{n} L_i$. In other words, the question is under which conditions
  \[
    \supN(K,L,P) = \|_{i=1}^{n} \supN(K_i,L_i,P^i_{i,o})\,.
  \]

  We show that the MOC condition plays a key role to answer this question. To prove the main result of this section, we make use of the following lemma.

  \begin{lem}\label{lemma_supNb}
    For a modular DES $G=\|_{i=1}^{n} G_i$ with $L=L(G)$ and $L_{m}=L_m(G)$. If $L$ is MOC with respect to projections $P_i$, $P$, and $P^{i}_{i.o}$, then normality of $S\subseteq L_m$ with respect to $L$ and $P$ implies normality of $P_i(S)$ with respect to projections $P_i(L)$ and $P^{i}_{i,o}$, for $i\in\{1,\ldots,n\}$.
  \end{lem}
  \begin{IEEEproof}
    We show that for any set $S\subseteq L_m$ that is normal with respect to $L$ and $P$, the set $P_i(S)$ is normal with respect to $P_i(L)$ and $P^i_{i,o}$, that is,
    $
      P_i(\overline{S}) = (P^i_{i,o})^{-1} P^i_{i,o} (P_i(\overline{S})) \cap P_i(L)
    $.

    However, since $P_i(\overline{S}) \subseteq (P^i_{i,o})^{-1} P^i_{i,o} (P_i(\overline{S})) \cap P_i(L)$, we need to show the opposite inclusion.
     To this end, we consider a string $t'\in (P^i_{i,o})^{-1}P^i_{i,o}(P_i(\overline{S})) \cap P_i(L)$. Then, there is a string $s\in \overline{S}$ such that $t' \in (P^i_{i,o})^{-1}P^i_{i,o}(P_i(s))$, and therefore $P^i_{i,o}(P_i(s)) = P^i_{i,o}(t')$. By the MOC property, there is a string $s'\in L$ such that $P_i(s')=t'$ and $P(s)=P(s')$, and hence
     \[
      s'\in P^{-1}P(s)\cap L \subseteq P^{-1}P(\overline{S}) \cap L = \overline{S}
    \]
    where the last equality is by normality of the set $S$. Altogether, we have shown that $t'=P_i(s')\in P_i(\overline{S})$, which was to be shown.
  \end{IEEEproof}

  To avoid the well-known conflicting issues and to focus on the role of MOC in maximal permissiveness, we consider prefix-closed languages or assume that the languages are nonconflicting. We point out that non-prefix-closed or conflicting languages can be handled by abstractions or coordinators for nonblockingness~\cite{JDEDS15,Pena2009,Malik2020}.

  We now formulate the main result of this section.
  \begin{thm}\label{thm15}
    Consider languages $K_i \subseteq L_i$ over $\Sigma_i$, where $L_i$ is prefix-closed, for $i=1,\ldots,n$ and $n\ge 2$. We define the global languages $K=\|_{i=1}^{n} K_i$ and $L = \|_{i=1}^{n} L_i$.
    \begin{outline}
      \1[1)] If the languages $\supN(K_i, L_i, P_{i,o}^i)$ are nonconflicting, then
      $
        \|_{i=1}^n \supN(K_i, L_i, P_{i,o}^i) \subseteq \supN(K,L,P).
      $

      \1[2)] If, in addition, for all $i=1,\ldots,n$,
        $P_i(L)=L_i$ and
        the language $L$ is MOC with respect to projections $P_{i}$, $P$, and $P_{i,o}^i$,
      then
      $
        \supN(K,L,P) = \|_{i=1}^n \supN(K_i, L_i, P_{i,o}^i).
      $
    \end{outline}
  \end{thm}
  \begin{IEEEproof}
    To simplify the notation, we denote the language $\supN(K_i, L_i, P_{i,o}^i)$ by $\supN_{i}$.

    To prove~1), we show that $\|_{i=1}^n \supN_{i}$ is normal with respect to $L$ and $P$, which implies that $\|_{i=1}^n \supN_{i}$ is included in the supremal element $\supN(K,L,P)$. To this end, we use the nonconflictingness of $\supN_i$, and we obtain that
    \begin{align*}
      \overline{\|_{i=1}^{n} \supN_i} 
      & \subseteq P^{-1}P(\overline{\|_{i=1}^n \supN_{i}}) \cap L \tag{by the properties of projections}\\
      & = P^{-1}P(\|_{i=1}^n \overline{\supN_{i}}) \cap L \tag{by the nonconflictingness of $\supN_i$}\\
      & \subseteq P^{-1}(\|_{i=1}^n  P_{i,o}^i(\overline{\supN_{i}})) \cap L \tag{by the projection of a parallel composition}\\
      & = \|_{i=1}^n ( P_{i,o}^i)^{-1} P_{i,o}^i(\overline{\supN_{i}}) \cap L \tag{by the properties of inverse projections}\\
      & = \|_{i=1}^n ( P_{i,o}^i)^{-1} P_{i,o}^i(\overline{\supN_{i}}) \cap \|_{i=1}^{n} L_i \tag{by replacing $L$ with $\|_{i=1}^{n} L_i$}\\
      & = \|_{i=1}^n [( P_{i,o}^i)^{-1} P_{i,o}^i (\overline{\supN_{i}}) \cap L_i] \tag{by reformulating the composition}\\
      & = \|_{i=1}^{n} \overline{\supN_i} \tag{by the normality of $\supN_i$}\\
      & = \overline{\|_{i=1}^{n} \supN_i} \tag{by the nonconflictingness of $\supN_i$}
    \end{align*}
    which shows that $\|_{i=1}^{n} \supN_i$ is normal, as claimed.
    
    To prove 2), the assumptions that $P_i(L)=L_i$ and that $L$ is MOC with respect to projections $P_i$, $P$, and $ P_{i,o}^i$, together with Lemma~\ref{lemma_supNb}, imply that the language $P_{i}(\supN(K, L, P))$ is normal with respect to $P_i(L)=L_i$ and $P_{i,o}^i$. As a result, we have that $P_{i}(\supN(K, L, P)) \subseteq \supN_{i}$, for all $i=1,\ldots,n$, which proves the claim.
  \end{IEEEproof}

\section{Ensuring MOC}\label{section7}
  Looking at Theorem~\ref{thm15}, the reader may see two key assumptions:
  (i) $P_i(L)=L_i$ and
  (ii) $L$ is MOC with respect to projections $P_{i}$, $P$, and $P_{i,o}^i$, for all $i=1,\ldots, n$.

  Whereas the condition $P_i(L)=L_i$ can be algorithmically verified, though in exponential time, it is in fact a modelling decision whether the condition $P_i(L)=L_i$ is satisfied. Indeed, $$L = \|_{i=1}^{n} L_i = \|_{i=1}^{n} P_i(L)\,,$$ and hence $P_i(L)$ can be considered instead of $L_i$; in addition, the technique described in Section~\ref{subsectionCMDES} handling the case of global specifications satisfies this condition by construction.

  The verification of MOC, on the other hand, is a task that we are currently unable to perform algorithmically. More specifically, the verification of MOC is a \PSpace-hard problem and it is open whether the problem is decidable.
  Therefore, the main algorithmic issue is how to (easily) verify or ensure that a given plant language satisfies MOC. In other words, the existence of conditions under which the MOC condition is satisfied is a challenging open problem.

  We now discuss this problem and show that the assumption that all shared events are observable is such a condition.

  To prove this result, we use the following well-known fact~\cite{Won04}. We also simplify the notation by denoting the languages $L(G_i)$ of local plants simply by $L_i$, for $i=1,\dots,n$.

  \begin{lem}\label{lemma:Wonham}
    Let $R\colon (\cup_{i=1}^{n} \Sigma_i)^* \to \Gamma^*$ be a projection, and let $R_i\colon \Sigma_i^* \to (\Gamma\cap\Sigma_i)^*$ be its restriction to local modules. If all shared events of the languages $L_i$ over $\Sigma_i$, for $i=1,\ldots,n$, are in $\Gamma$, that is, $\Sigma_{s}\subseteq \Gamma$, then $R(\|_{i=1}^n L_i) = \|_{i=1}^n R_i(L_i)$.
    \hfill\IEEEQEDhere
  \end{lem}

  We can now formulate the following key lemma showing that the easily-verifiable assumption that all shared event are observable is of interest in supervisory control of modular systems. It is worth noticing that this assumption is similar to the frequently used assumption that all shared events are controllable, used in supervisory control of modular systems under complete observation~\cite{Gaudin2007,WILLNER1991}.

 \begin{lem}\label{conjecture_proven}\label{lemma9c}
    Given a modular plant $L = \|_{i=1}^{n} L_i$ that is formed by prefix-closed languages $L_i$ over $\Sigma_i$, for $i=1,\ldots,n$. If all shared events are observable, then the language $L$ is MOC with respect to projections $P_{i}$, $P$, and $P_{i,o}^i$, for all $i=1,\ldots,n$.
  \end{lem}
  \begin{IEEEproof}
    To show that $L$ is MOC with respect to $P_i$, $P$, and $ P_{i,o}^i$, for all $i=1,\ldots,n$, let $s\in L$ and $t_i \in P_i(L)\subseteq L_i$ be two arbitrary strings such that $P_{i,o}^i P_i (s) =  P_{i,o}^i (t_i)$. We need to show that there is a string $s'\in L$ such that $P(s')=P(s)$ and $P_i(s')=t_i$. To this end, we first notice that
    \[
      P(s) \parallel t_i \parallel \|_{j\neq i} P_{j}(s) \neq \emptyset
    \]
    if and only if
    \[
      P\bigl(P(s) \parallel t_i \parallel \|_{j\neq i} P_{j}(s)\bigr) \neq \emptyset\,.
    \]

    However, because all shared events are observable, that is, $\Sigma_s \subseteq \Sigma_o$, we have that
    \begin{align*}
      & P(P(s) \parallel t_i \parallel \|_{j\neq i} P_{j}(s)) \\
      & = P(s) \parallel P^i_{i,o}(t_i) \parallel \|_{j\neq i} P^j_{j,o}P_{j}(s) \tag{by Lemma~\ref{lemma:Wonham}}\\
      & = P(s) \parallel P^i_{i,o}P_i(s) \parallel \|_{j\neq i} P^j_{j,o}P_{j}(s) \tag{by the assumption}\\
      & = P(s) \parallel \|_{i=1}^{n} P^i_{i,o}P_{i}(s) \\
      & = P(s) \parallel \|_{i=1}^{n} P_{i,o}^o P(s) \tag{by commutativity of Figure~\ref{projections}}
    \end{align*}

    Since $P(s) \in P(s) \parallel \|_{i=1}^{n} P_{i,o}^o P(s)$, we have that $P(P(s) \parallel t_i \parallel \|_{j\neq i} P_{j}(s)) \neq \emptyset$, and hence $P(s) \parallel t_i \parallel \|_{j\neq i} P_{j}(s) \neq \emptyset$.

    Therefore, considering any
    $
      s' \in P(s) \parallel t_i \parallel \|_{j\neq i} P_{j}(s)
      \subseteq P(L) \parallel L_i \parallel \|_{j\neq i} L_j
      = P(L) \parallel L
       = L
    $,
    we have that the string $s'\in L$, $P(s')=P(s)$, and $P_i(s')=t_i$, as required.
  \end{IEEEproof}

  We now formulate a result for prefix-closed specifications.
  \begin{cor}\label{cor01}
    Given prefix-closed languages $L_i$ over $\Sigma_i$, for $i=1,\ldots,n$ with $n\ge 2$. If the alphabets $\Sigma_1,\ldots,\Sigma_n$ are pairwise disjoint, then, for all $i=1,\ldots,n$, $P_i(L)=L_i$ and the language $L = \|_{j=1}^{n} L_j$ is MOC with respect to projections $P_{i}$, $P$, and $P_{i,o}^i$.
  \end{cor}
  \begin{IEEEproof}
    Since the alphabets are pairwise disjoint, the set of shared events $\Sigma_s=\emptyset \subseteq \Sigma_i$, for all $i$, and Lemma~\ref{lemma:Wonham} implies that $P_i(L)=P_i(\|_{j=1}^{n} L_j) = \|_{j=1}^{n} P_i(L_j) = L_i$, as claimed.

    The rest follows from Lemma~\ref{lemma9c}.
  \end{IEEEproof}

  \begin{remark}
    It is worth pointing out that the result of Komenda and Van Schuppen~\cite[Theorem~5.13]{KvS_TCS07}, which shows that $$\supN(K,L,P) = \|_{i=1}^n \supN(K_i, L_i, P_{i,o}^i)$$ for prefix-closed local specifications over disjoint alphabets of local modules, is a consequence of Corollary~\ref{cor01} and Theorem~\ref{thm15}.
  \end{remark}

  We are now ready to present the main result of this section.
  \begin{thm}\label{cor_conjecture_proven}
    Let $n\ge 2$, and let $L = \|_{i=1}^{n} L_i$ be a modular DES formed by prefix-closed languages $L_i$ over $\Sigma_i$. Let $K = \|_{i=1}^{n} K_i$ with $K_i\subseteq L_i$ be a decomposable specification. If the languages $\supN(K_i, L_i, P_{i,o}^i)$ are nonconflicting, $P_i(L)=L_i$, and all shared events are observable, then
    \[
      \supN(K,L,P) = \|_{i=1}^n \supN(K_i, L_i, P_{i,o}^i)\,.
    \]
  \end{thm}
  \begin{IEEEproof}
    Since $\Sigma_s \subseteq \Sigma_o$, Lemma~\ref{conjecture_proven} implies that $L$ is MOC with respect to projections $P_{i}$, $P$, and $P_{i,o}^i$, for all $i=1,\ldots,n$. The result then follows by the application of Lemma~\ref{conjecture_proven} and Theorem~\ref{thm15}.
  \end{IEEEproof}

  Finally, we point out that neither $\Sigma_o \subseteq \Sigma_{s}$ nor $\cap_j \Sigma_{j}\subseteq \Sigma_o$ is a suitable condition ensuring MOC. Indeed, for the languages
  $L_1=\overline{\{u_1uc\}} \subseteq \{u_1,u,c\}^*$, $L_2=\overline{\{u_2cu\}}\subseteq \{u_2,c,u\}^*$, and $L_3=\overline{\{u_3c\}}\subseteq \{u_3,c\}^*$,
  and their respective specifications
    $K_1=\overline{\{u_1\}}$,
    $K_2=\overline{\{u_2\}}$, and
    $K_3=\overline{\{u_3\}}$,
  where $\Sigma_{1,o}=\Sigma_{2,o}=\Sigma_{3,o}=\{c\}$, and hence $\Sigma_o=\{c\}$, we obtain that
  $\supN(K_1,L_1,P^1_{1,o})=\emptyset$,
  $\supN(K_2,L_2,P^2_{2,o})=\{u_2\}$, and
  $\supN(K_3,L_3,P^3_{3,o})=\{u_3\}$,
  while
  $
    \supN(K_1\|K_2\|K_3,L_1\|L_2\|L_3,P)=K_1\parallel K_2\parallel K_3.
  $
  In particular,
  $
    P_1(\supN(K_1\|K_2\|K_3,L_1\|L_2\|L_3,P)) = u_1 \notin \supN(K_1,L_1,P^1_{1,o}) = \emptyset.
  $
  This example provides a counterexample for both cases, because $\{c\} = \Sigma_o \subseteq \Sigma_s = \{c,u\}$ and $\{c\} = \Sigma_1 \cap \Sigma_2 \cap \Sigma_3 \subseteq \Sigma_o = \{c\}$.

\begin{figure}
    \centering
    \includegraphics[scale=.94]{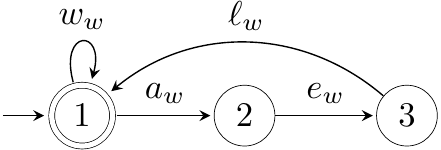}\quad
    \includegraphics[scale=.94]{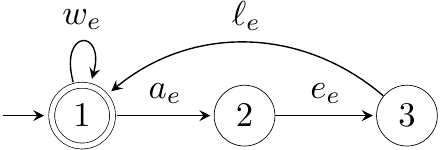}
    \caption{Generators $G_1$ and $G_2$ modelling the trains $T_1$ and $T_2$, resp.}\label{figGens}
  \end{figure}

  We now provide an illustrative example.
  \begin{exmp}\label{ex1}
    We consider the synthesis of a bridge controller of a railroad with two tracks and a bridge where the tracks merge motivated by Alur~\cite{Alur}.
    Two trains operate in the system---the western train $T_1$ and the eastern train $T_2$. To control the access to the bridge, trains communicate with the bridge controller. If the western train arrives at the bridge, it sends the arrive signal $a_w$. If the bridge controller accepts the signal, the train can enter the bridge ($e_w$); otherwise, it waits ($w_w$) and keeps sending the arrive signal $a_w$ until it is accepted. When leaving the bridge, the train sends the leave signal $\ell_w$. The eastern train behaves the same, using the signals $a_e$, $e_e$, $w_e$, and $\ell_e$, respectively.
    The models $G_1$ and $G_2$ of the two trains are depicted in Figure~\ref{figGens}. To simplify the notation, we define $L_i = L(G_i)$, for $i=1,2$, and $L= L_1 \parallel L_2$.

    We primarily focus on partial observation, and therefore we assume that all events are controllable. The observable events are $\Sigma_o=\{w_w, w_e, a_w, a_w, e_e, e_w\}$.

    Since the alphabets of local modules are disjoint, Corollary~\ref{cor01} gives that $P_1(L)=L_1$ and $P_2(L)=L_2$, and that the language $L=L_1\|L_2$ is MOC with respect to projections $P_i$, $P$, and $P_{i,o}^i$, for $i=1,2$.
    However, these results are the assumptions of Theorem~\ref{thm15}, and hence the theorem is applicable to any local specifications $K_i$, for $i=1,2$, for which the supremal normal languages $\supN(K_i,L_i, P_{i,o}^i)$ are nonconflicting.

    For an illustration, we consider a local safety specification that a train first waits, then its request is accepted, followed by entering and leaving the bridge:
    $K_1=(w_w a_w e_w \ell_w)^*$ and $K_2=(w_e a_e e_e \ell_e)^*$.
    The local supervisors realize the supremal normal languages $\supN(K_i,L_i,P_{i,o}^i)=K_i$ and the reader may verify that $\supN(K_1\|K_2,L,P) = K_1 \| K_2$.

    We point out that this specification is only illustrative and that we discuss a realistic specification in Example~\ref{ex2}.
  \end{exmp}

\section{Synthesis of Normal Supervisors for Modular DES with Global Specifications}
\label{subsectionCMDES}
  In this section, we discuss the case of global specifications. The main idea is to use the concept of conditional decomposability of our coordination control framework, where we use it to handle the interaction among various local modules~\cite{KMvS2011}.

  Specifically, for a modular system with local languages $L_i$ over $\Sigma_i$, and a global specification $K\subseteq L = \|_{i=1}^{n} L_i$, we find an alphabet
  \[
    \Sigma_\kappa \supseteq \Sigma_s
  \]
  containing all shared events in such a way that the specification $K$ is {\em conditionally decomposable} with respect to the alphabets of local modules extended with the events of $\Sigma_\kappa$; formally,
  \[
    K = \|_{i=1}^{n} P_{i+\kappa}(K)
  \]
  where $P_{i+\kappa}\colon \Sigma^* \to \Sigma_{i+\kappa}^*$ is the projection from the global alphabet $\Sigma=\cup_{i=1}^{n} \Sigma_i$ to the local alphabet $\Sigma_{i+\kappa} = \Sigma_i \cup \Sigma_\kappa$ of the local module $L_i$ extended with the events of $\Sigma_\kappa$, see Komenda et al.~\cite{JDEDS15} for more details.

  The idea of the coordination approach is to decompose the global specification $K$ into local specifications $P_{i+\kappa}(K)$ over the local alphabets $\Sigma_{i+\kappa}$, for $i=1,\dots,n$; note that there is always a suitable alphabet $\Sigma_{\kappa}$, for which the specification $K$ is conditionally decomposable---if there were no better choice, then the choice of $\Sigma_{\kappa}=\Sigma$ would work.
  The verification of conditional decomposability as well as the computation of a suitable alphabet $\Sigma_\kappa$ is of  polynomial-time complexity, although the computation of the minimal $\Sigma_\kappa$ with respect to set inclusion is \NP-hard~\cite{SCL12}.

  Having determined an alphabet $\Sigma_\kappa$, we compute a coordinator $G_\kappa$ as a DFA satisfying
  \[
    L_\kappa = L(G_\kappa) = P_{\kappa}(L) = \|_{i=1}^{n} P^i_{\kappa,i}(L_i)
  \]
  where $P_\kappa\colon \Sigma^* \to \Sigma_\kappa^*$ and $P^i_{\kappa,i}\colon \Sigma_i^* \to \Sigma_{i,\kappa}^*$ are projections.

  This way, we have transformed the original modular system consisting of modules $L_i$ over $\Sigma_i$ and a global specification $K$ to a modular system consisting of modules
  \begin{align}\label{lik}
    L_{i+\kappa} & = P_{i+\kappa}(L) = L_i \parallel L_\kappa
  \end{align}
  and the local specifications
  \begin{align}\label{kik}
    K_{i+\kappa} = P_{i+\kappa}(K)
  \end{align} 
  for which the plant language is $L = \|_{i=1}^{n} L_i = \|_{i=1}^{n} L_{i+\kappa}$ and the specification is $K = \|_{i=1}^{n} K_{i+\kappa}$,
  see Komenda et al.~\cite{JDEDS15} for more details.
  
  \begin{figure}
    \centering
    \includegraphics[scale=1]{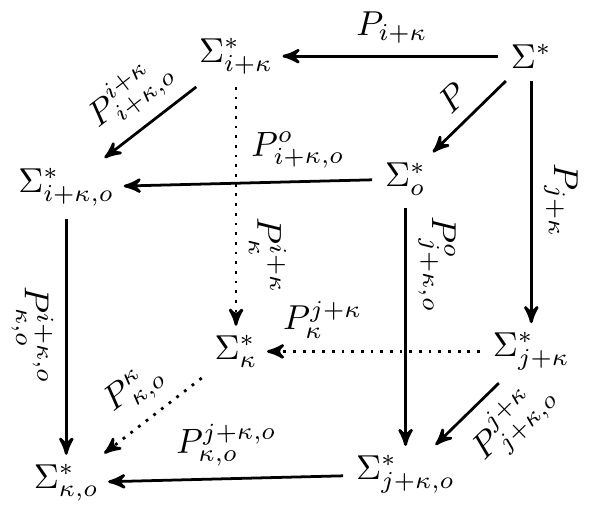}
    \caption{Projections used in the coordination control framework.}
    \label{fig1b}
  \end{figure}

  We now obtain the following corollary of Lemma~\ref{conjecture_proven}; see Figure~\ref{fig1b} for the notation of the used projections.

  \begin{lem}\label{cor_2n}
    Consider a modular DES $G = \|_{i=1}^{n} G_i$, where $\Sigma_s$ denotes the set of shared events. Let $L_i = L(G_i)$. Then, for every alphabet $\Sigma_\kappa\supseteq \Sigma_s$, whenever $\Sigma_{\kappa} \subseteq \Sigma_o$, the global plant $L = \|_{i=1}^n L_i$ is MOC with respect to projections $P_{j+\kappa}$, $P$, and $P^{j+\kappa}_{j+\kappa,o}$, for all $j=1,\dots, n$, where $P^{j+\kappa}_{j+\kappa,o}\colon \Sigma_{j+\kappa}^* \to \Sigma_{j+\kappa,o}^*$.
  \end{lem}
  \begin{IEEEproof}
    The proof follows from Lemma~\ref{conjecture_proven} applied to the language $L=\|_{i=1}^n L_{i+\kappa}=\|_{i=1}^n L_{i}$, because the shared events of $L_{i+\kappa}$, for $i=1,\ldots,n$, are observable---the set of shared events is the alphabet $\Sigma_\kappa \subseteq \Sigma_o$. See also Figure~\ref{fig1b} for the notation of projections.
  \end{IEEEproof}
  
  We can now state the main result of this section.
  \begin{thm}\label{mainCC}
    Consider prefix-closed languages $L_i$ over $\Sigma_i$, for $i=1,\ldots,n$, and define the global language $L = \|_{i=1}^{n} L_i$. Let $K \subseteq L$ be a global specification. Compute an alphabet $\Sigma_\kappa$ containing all shared events such that $\Sigma_\kappa$ makes the specification $K$ conditionally decomposable with respect to alphabets $\Sigma_{i+\kappa}$, $i=1,\ldots,n$. If, for the languages $L_{i+\kappa}$ and $K_{i+\kappa}$ defined in \eqref{lik} and \eqref{kik}, respectively,
    \begin{enumerate}
      \item languages $\supN(K_{i+\kappa}, L_{i+\kappa}, P_{i+\kappa ,o}^{i+\kappa})$ are nonconflicting, and
      \item $\Sigma_\kappa \subseteq \Sigma_o$ consists only of observable events,
    \end{enumerate}
    then
    $
      \supN(K,L,P) = \|_{i=1}^n \supN(K_{i+\kappa}, L_{i+\kappa},  P_{i+\kappa ,o}^{i+\kappa}).
    $
  \end{thm}
  \begin{IEEEproof}
    The result follows from Lemma~\ref{cor_2n} and Theorem~\ref{cor_conjecture_proven},
    because $L_{i+\kappa} = P_{i+\kappa}(L)$, for $i=1,\ldots,n$, by definition, and the alphabet $\Sigma_\kappa$ forms the set of shared events of local modules $L_{i+\kappa}$, for $i=1,\ldots,n$.
  \end{IEEEproof}
  
  We now proceed with the railroad example, discussing a realistic global specification.
  \begin{exmp}\label{ex2}
    We again consider the synthesis of a bridge controller described in Example~\ref{ex1}, where the train models are depicted in Figure~\ref{figGens}, and we post the safety requirement that a train may enter the bridge if its arrive signal is accepted. The signal may be accepted if the other train waits at or is away from the bridge and there is no train on the bridge.

    Our specification depicted in Figure~\ref{figSpec1} takes care of this requirement as well as of a kind of fairness; namely, both trains wait before the arrive signal of one of them is accepted, and no train that wants to enter the bridge waits for ever.
    \begin{figure}
      \centering
      \includegraphics[scale=1,align=c]{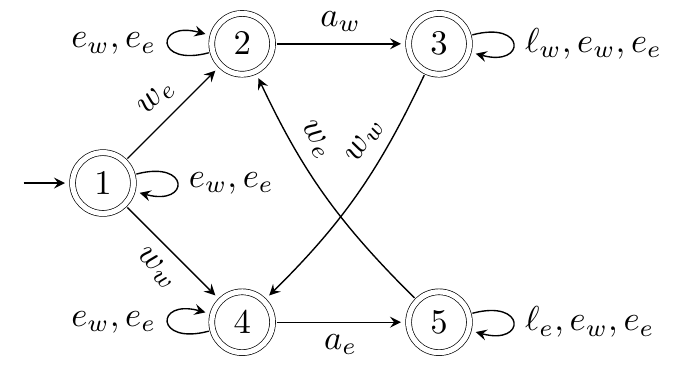}
      \caption{The global specification $K$.}\label{figSpec1}
    \end{figure}
    We recall that unobservable events are $\Sigma_{uo} = \{\ell_w, \ell_e\}$.

    Since the specification is global, we look for an alphabet $\Sigma_\kappa$ that contains all shared events and that makes our specification, which we denote by $K$, conditionally decomposable.
    The alphabets of local modules are $\Sigma_1=\{a_w,e_w,\ell_w,w_w\}$ and $\Sigma_2=\{a_e,e_e,\ell_e,w_e\}$, and hence the set of shared events is $\Sigma_s=\emptyset$.
    The reader may verify that the alphabet $$\Sigma_\kappa=\{w_w,w_e\}$$ makes the specification $K$ of Figure~\ref{figSpec1} conditionally decomposable, that is,
    $
      K = P_{1+\kappa}(K) \parallel P_{2+\kappa} (K).
    $
    We now compute the language
    \[
      L_\kappa = \|_{i=1}^{n} P_{\kappa,i}^i(L_i) = \{r_1,r_2\}^*
    \]
    and the new modules $L_{i+\kappa} = L_i \parallel L_\kappa$, see Figure~\ref{Lik}.
    \begin{figure}
      \centering
      \includegraphics[scale=1]{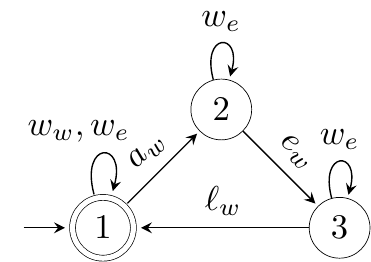}\quad
      \includegraphics[scale=1]{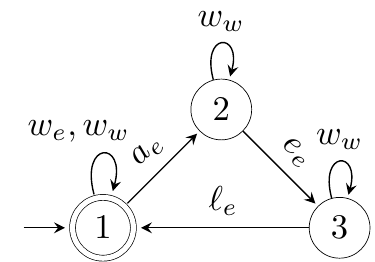}
      \caption{New modules $L_{1+\kappa}$ and $L_{2+\kappa}$.}
      \label{Lik}
    \end{figure}

    Having decomposed the global specification $K$ into local specifications $K_{i+\kappa} = P_{i+\kappa}(K)$, as depicted in Figure~\ref{Kik},
    \begin{figure}
      \centering
      \includegraphics[scale=.92]{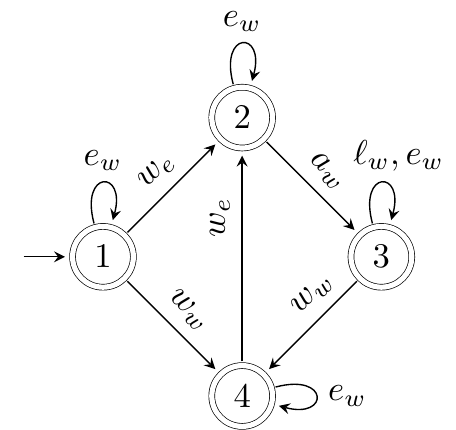}\quad
      \includegraphics[scale=.92]{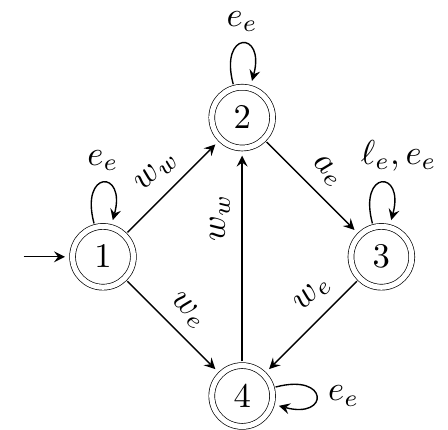}
      \caption{Decomposition of the global specification $K$ into local specifications $K_{i+\kappa} = P_{i+\kappa}(K)$ of the new modules $L_{i+\kappa}$.}
      \label{Kik}
    \end{figure}
    we proceed by computing the local normal supervisors realizing the supremal normal languages
    $
      \supN(K_{i+\kappa},L_{i+\kappa}, P_{i+\kappa ,o}^{i+\kappa})
    $,
    for $i=1,2$, depicted in Figure~\ref{supNi_global}.
    \begin{figure}
      \centering
      \includegraphics[scale=1]{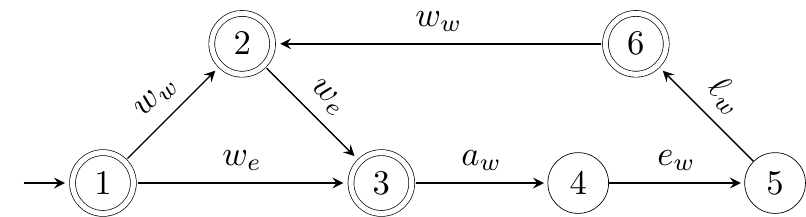}\quad
      \includegraphics[scale=1]{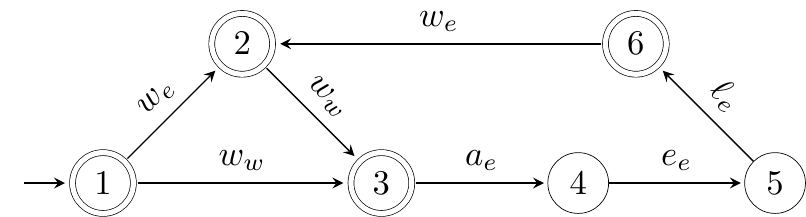}
      \caption{The constructed local normal supervisors.}
      \label{supNi_global}
    \end{figure}

    Since the alphabet $\Sigma_\kappa\subseteq \Sigma_o$ consists only of observable events, and the languages $\supN(K_{i+\kappa},L_{i+\kappa}, P_{i+\kappa ,o}^{i+\kappa})$ are prefix-closed, and hence nonconflicting, Theorem~\ref{mainCC} gives that
    \[
      \supN(K,L,P) = \|_{i=1}^{n} \supN(K_{i+\kappa},L_{i+\kappa},P_{i+\kappa,o}^{i+\kappa})\,.
    \]
  \end{exmp}

\section{Modular Computation of Supremal Controllable and Normal Sublanguages}\label{consequences}
  Under complete observation, Zhong and Wonham~\cite{ZhongW1990} formulated the notion of {\em output control consistency\/} (OCC) that together with the {\em observer property\/} guarantee that controllability is preserved under projections.

  For two alphabets $\Gamma \subseteq \Sigma$, the projection $R\colon \Sigma^* \to \Gamma^*$ is {\em output control consistent\/} (OCC) for a prefix-closed language $L$ over $\Sigma$ if for every string $s\in L$ of the form $s=\sigma_1\cdots \sigma_k$ or $s = s' \sigma_1 \cdots \sigma_k$, $k\ge 1$, satisfying that (i) the prefix $s'$ terminates with an event from $\Gamma$, (ii) the event $\sigma_i \in \Sigma - \Gamma$, for $i=1,\ldots,k-1$, and (iii) the last event $\sigma_k \in \Gamma$, we have that if the event $\sigma_k$ is uncontrollable, then so are uncontrollable events $\sigma_i$, for $i=1,\ldots,k-1$.

  The projection $R\colon \Sigma^* \to \Gamma^*$ is an {\em $L_m(G)$-observer} for a nonblocking plant $G$ over $\Sigma$ if for all strings $t\in R(L_m(G))$ and $s\in \overline{L_m(G)}$, whenever the projection $R(s)$ is a prefix of $t$, then there is $u\in \Sigma^*$ such that $su\in L_m(G)$ and $R(su)=t$.

  We now recall the result of Feng~\cite{FLT} that OCC and the observer property guarantee that controllability is preserved by projections. This result is a complete-observation counterpart of Lemma~\ref{lemma_supNb}.
  \begin{lem}[{\cite[Lemma~4.3]{FLT}}]
  \label{preservation_controllability}
    Let $L$ and $X$ be prefix-closed languages over an alphabet $\Sigma$, such that $X\subseteq L$. Suppose that a language $S\subseteq X$ is controllable with respect to $L$ and $\Sigma_{uc}$. If the projection $R\colon \Sigma^* \to \Gamma^*$ is an $X$-observer and OCC for $L$, then $R(S)$ is controllable with respect to $R(X)$ and $\Sigma_{uc}\cap \Gamma$.\hfill\IEEEQED
  \end{lem}

  Schmidt and Breindl~\cite{SB11} further investigated the problem of Lemma~\ref{preservation_controllability} and defined a weaker version of the OCC condition, called {\em local control consistency} (LCC). The projection $R\colon \Sigma^* \to \Gamma^*$ is {\em locally control consistent\/} (LCC) for a string $s\in L(G)$ if for every $e \in \Gamma \cap \Sigma_{uc}$ with $R(s)e \in R(L(G))$ either there is no $u\in (\Sigma - \Gamma)^*$ such that $sue \in L(G)$ or there is $u\in (\Sigma_{uc} - \Gamma)^*$ such that $sue\in L(G)$. We say that the projection $R$ is LCC for a language if it is LCC for all strings of the language.

  \begin{lem}[{\cite[Lemma~4.1]{SB11}}]
  \label{controllability}
    Let $G$ over $\Sigma$ be a nonblocking plant, let $T$ over $\Gamma\subseteq \Sigma$ be a specification, and let $R\colon \Sigma^* \to \Gamma^*$ be the corresponding projection. If $\supC$ is the supremal controllable sublanguage of $T\,\|\,L_m(G)$ with respect to $L(G)$ and $\Sigma_{uc}$, and the projection $R$ is an $L_m(G)$-observer and LCC for $L(G)$, then the language $R(\supC)$ is controllable with respect to $R(L(G))$ and $\Sigma_{uc}\cap \Gamma$.\hfill\IEEEQED
  \end{lem}

  Because the proof of Lemma~\ref{controllability} does not depend on the supremality of the considered controllable language, we can reformulate it as follows.

  \begin{lem}
  \label{pres_controllability}
    Let $L$ over $\Sigma$ be a prefix-closed language, and suppose that a language $S\subseteq  L$ is controllable with respect to $L$ and $\Sigma_{uc}$. If the projection $R\colon \Sigma^* \to \Gamma^*$ is an $L$-observer and LCC for $L$, then $R(S)$ is controllable with respect to $R(L)$ and $\Sigma_{uc}\cap \Gamma$.\hfill\IEEEQED
  \end{lem}

  We now combine our results for normality with the existing results for controllability. We use the notation $$\supCN(K,L,\Sigma_{uc},P)$$ to denote the supremal controllable and normal sublanguage of the specification $K$ with respect to the plant language $L$, the set of uncontrollable events $\Sigma_{uc}$, and the partial-observation projection $P\colon \Sigma^*\to\Sigma_o^*$.

  \begin{thm}\label{thm15CN}
    For an integer $n\ge 2$, we consider a modular DES $L = \|_{i=1}^{n} L_i$ formed by prefix-closed languages $L_i$ over $\Sigma_i$, and a decomposable specification $K = \|_{i=1}^{n} K_i$, where $K_i\subseteq L_i$. If, for all $i=1,\dots,n$,
    \begin{enumerate}
      \item $\supCN(K_i, L_i,\Sigma_{i,uc},P_{i,o}^i)$ are nonconflicting,
      \item $P_i(L)=L_i$,
      \item the projection $P_i\colon \Sigma^* \to \Sigma_i^*$ is an $L$-observer,
      \item $P_i\colon \Sigma^* \to \Sigma_i^*$ is LCC (or OCC) for $L$, and
      \item $L$ is MOC with respect to projections $P_{i}$, $P$, and $P_{i,o}^i$,
    \end{enumerate}
    then
    \[
      \supCN(K,L,\Sigma_{uc},P) = \|_{i=1}^n \supCN(K_i, L_i, \Sigma_{i,uc}, P_{i,o}^i)\,.
    \]
  \end{thm}
  \begin{IEEEproof}
    The preservation of normality can be shown analogously as in the proof of Theorem~\ref{thm15}.

    For controllability, the right-to-left inclusion ``$\supseteq$'' follows from Proposition~4.6 of Feng~\cite{FLT} showing that the parallel composition of nonconflicting controllable languages is controllable. The left-to-right inclusion follows from Lemma~\ref{pres_controllability} (resp. Lemma~\ref{preservation_controllability}) by substituting $P_i$ for $R$, for $i=1,\ldots,n$, which shows that the language $P_i(\supCN(K,L,\Sigma_{uc},P))\subseteq K_i$ is controllable with respect to $P_i(L)=L_i$ and $\Sigma_{i,uc}$, and hence it is included in $\supCN(K_i, L_i, \Sigma_{i,uc}, P_{i,o}^i)$.
  \end{IEEEproof}

  For completely observed modular DES with prefix-closed specifications, Willner and Heymann~\cite[Theorem~4.4]{WILLNER1991} have shown that if all shared events are controllable, then the nonblocking and maximally permissive supervisor can be computed in a modular way.
  This result can be lifted to non-prefix-closed specifications with the help of Proposition 4.6 of Feng~\cite{FLT} as discussed in the proof of Theorem~\ref{thm15CN}. We further lift it to partially observed modular systems as follows.

  \begin{cor}\label{cor17}
    For an integer $n\ge 2$, we consider a modular DES $L = \|_{i=1}^{n} L_i$, formed by prefix-closed languages $L_i$ over $\Sigma_i$, and a decomposable specification $K = \|_{i=1}^{n} K_i$, where $K_i\subseteq L_i$. If, for all $i=1,\ldots,n$,
    \begin{enumerate}
      \item $\supCN(K_i, L_i, \Sigma_{i,uc}, P_{i,o}^i)$ are nonconflicting,
      \item $P_i(L)=L_i$, and
      \item all shared events are controllable and observable,
    \end{enumerate}
    then
    \[
      \supCN(K,L,\Sigma_{uc},P) =  \|_{i=1}^n \supCN(K_i, L_i, \Sigma_{i,uc}, P_{i,o}^i)\,.\IEEEQEDhereeqn
    \]
  \end{cor}

  Combining Theorem~\ref{thm15CN} and Corollary~\ref{cor17} with the reduction of Section~\ref{subsectionCMDES}, transforming a modular DES with a global specification to a modular DES with local specifications, we obtain the following results.

  \begin{thm}\label{thm15CNglob}
    For an integer $n\ge 2$, we consider a modular DES $L = \|_{i=1}^{n} L_i$ formed by prefix-closed languages $L_i$ over $\Sigma_i$, and a global specification $K \subseteq L$.
    Compute an alphabet $\Sigma_\kappa$ containing all shared events such that $\Sigma_\kappa$ makes the specification $K$ conditionally decomposable with respect to alphabets $\Sigma_{i+\kappa}$, for $i=1,\ldots,n$.
    If, for the languages $L_{i+\kappa}$ and $K_{i+\kappa}$ defined in \eqref{lik} and \eqref{kik}, respectively,
    \begin{enumerate}
      \item the languages $\supCN(K_{i+\kappa}, L_{i+\kappa}, \Sigma_{i+\kappa,uc}, P_{i+\kappa ,o}^{i+\kappa})$ are nonconflicting,
      \item the projection $P_{i+\kappa}\colon \Sigma^* \to \Sigma_{i+\kappa}^*$ is an $L$-observer,
      \item $P_{i+\kappa}\colon \Sigma^* \to \Sigma_{i+\kappa}^*$ is LCC (or OCC) for $L$, and
      \item the language $L$ is MOC with respect to projections $P_{i+\kappa}$, $P$, and $P^{i+\kappa}_{i+\kappa,o}$,
    \end{enumerate}
    for all $i=1,\dots, n$, then
    \begin{multline*}
      \supCN(K,L,\Sigma_{uc},P)\\ =  \|_{i=1}^n \supCN(K_{i+\kappa}, L_{i+\kappa}, \Sigma_{i+\kappa,uc}, P_{i+\kappa,o}^{i+\kappa}).
    \end{multline*}
  \end{thm}
  \begin{IEEEproof}
    A proof follows by the application of Theorem~\ref{thm15CN} to a modular DES with local specifications constructed from a modular DES with a global specification in Section~\ref{subsectionCMDES}.
  \end{IEEEproof}

  We can now formulate the following corollary.
  \begin{cor}\label{cor19}
    For an integer $n\ge 2$, we consider a modular DES $L = \|_{i=1}^{n} L_i$, formed by prefix-closed languages $L_i$ over $\Sigma_i$, and a global specification $K \subseteq L$.
    We compute an alphabet $\Sigma_\kappa$ containing all shared events such that $\Sigma_\kappa$ makes $K$ conditionally decomposable with respect to $\Sigma_{i+\kappa}$, for $i=1,\ldots,n$.
    If, for the languages $L_{i+\kappa}$ and $K_{i+\kappa}$ defined in \eqref{lik} and \eqref{kik}, respectively, $\supCN(K_{i+\kappa}, L_{i+\kappa}, \Sigma_{i+\kappa,uc}, P_{i+\kappa ,o}^{i+\kappa})$ are nonconflicting, and either
    \begin{itemize}
      \item for $i=1,\ldots,n$, $P_{i+\kappa}\colon \Sigma^* \to \Sigma_{i+\kappa}^*$ is an $L$-observer and LCC (or OCC) for $L$, and the alphabet $\Sigma_\kappa$ contains only observable events,
    \end{itemize}
    or
    \begin{itemize}
      \item $\Sigma_\kappa$ contains only controllable and observable events,
    \end{itemize}
    then
    \begin{multline*}
      \supCN(K,L,\Sigma_{uc},P)\\ =  \|_{i=1}^n \supCN(K_{i+\kappa}, L_{i+\kappa}, \Sigma_{i+\kappa,uc}, P_{i+\kappa,o}^{i+\kappa})\,.
    \end{multline*}
    \hfill\IEEEQEDhere
  \end{cor}

  In a similar way, we could combine our results with the results of De Queiroz and Cury~\cite{Queiroz2000}, Hill and Tilbury~\cite{HillTilbury2006}, or Gaudin and Marchand~\cite{Gaudin2007}.

\section{Case Study}\label{section8}
  To evaluate our results on an industrial example, we consider the model and specification of a patient table of an MRI scanner designed by Theunissen~\cite{theunissen}. The plant consists of four components
  \[
    \text{VAxis}  \parallel  \text{HAxis} \parallel  \text{HVN}  \parallel \text{UI}\,.
  \]
  Although each component is again a composition of other components, we do not go into more details and consider these four modules as the modular DES. Similarly, the specification consists of four parts
  \[
    \text{VReq}  \parallel  \text{HReq}  \parallel  \text{HVReq}  \parallel \text{UIReq}\,,
  \]
  which do not exactly correspond the the four modules. In fact, each of the four parts of the specification concerns some of the four modules, and therefore each part of the specification can be seen as a global specification for the concerned modules.

  The model of Theunissen~\cite{theunissen} is constructed under complete observation. To lift it to partial observation, we define the events occurring in the specification as observable, and the other as unobservable.
  However, since we are currently unable to algorithmically verify MOC, and the models are too large for a manual verification, we need to ensure the MOC condition by making all shared (resp. coordinated) events observable, as required in Theorems~\ref{cor_conjecture_proven} and~\ref{mainCC}.

  For the computations, we used the C++ library \texttt{libFAUDES} in version 2.31d~\cite{libfaudes}. The computations were performed on an Inter-Core i7 processor laptop with 15 GB memory running Ubuntu 22.04.

  We computed the automata representations of the supervisors using the \texttt{libFAUDES} function \texttt{SupConNormNB}, which implements the standard algorithm for the computation of the supremal controllable and normal sublanguage. The automata are further minimized with respect to the number of states using the function \texttt{StateMin}.

  \subsection{Procedure}
  We now describe the procedure how we handle the computation of local supervisors. Since we consider every part of the specification as a global specification, our procedure is based on the coordination approach described in Section~\ref{subsectionCMDES}.
  \begin{outline}
    \1 For every $K\in\{\text{VReq},\text{HReq},\text{HVReq},\text{UIReq}\}$,
    we take the set $H$ of all plants from $\{\text{VAxis},\text{HAxis},\text{HVN},\text{UI}\}$ that share an event with $K$.

    \1 Let the alphabet of $K$ be denoted by $\Sigma_K$. If $\Sigma_K$ is strictly included in the set of events occurring in the plants of $H$, which we denote by $\Sigma_H$, we lift the specification $K$ to the alphabet $\Sigma_H$ by the inverse of projection $R\colon \Sigma_H^* \to \Sigma_K^*$.
      \2 Now, the specification is $R^{-1}(K)$.

    \1 If needed, we make the specification $R^{-1}(K)$ conditionally decomposable with respect to the alphabets of the modules of $H$ by computing an alphabet $\Sigma_{\kappa}$ using the \texttt{libFAUDES} function \texttt{ConDecExtension} from the coordination control plug-in.
      \2 The set of observable events is $\Sigma_o = \Sigma_K \cup \Sigma_{\kappa}$.
      \2 Controllable events are defined by Theunissen~\cite{theunissen}.

    \1 We consider the pair $(H,R^{-1}(K))$ as an instance of the modular DES consisting of the plans of $H$ and of the global specification $R^{-1}(K)$.
      \2 We use Corollary~\ref{cor19} to construct local supervisors. The constructed supervisors are then maximally permissive in the monolithic sense.
      \2 Although we do not discuss the conditions of Corollary~\ref{cor19} for controllability below, we mention here that they are also satisfied.
  \end{outline}

  \subsection{Results}
  In this section, we briefly discuss the obtained results, which we summarize in the corresponding tables.

  The specification VReq contains nine events that are shared only with the plant VAxis. Therefore, $H=\{\text{VAxis}\}$. The events of VReq are the only observable events. The automaton representation of VReq has 12 states and 44 transitions, while the representation of VAxis has 15 states and 50 transitions. A monolithic approach  was used to compute a supervisor $S$ with 15 states and 36 transitions.
  The results for the specifications VReq are summarized in Table~\ref{tabVAxisA}, where $S$ denotes the automaton representing the controllable and normal supervisor realizing the supremal controllable and normal sublanguage of the specification.

  \begin{table}[ht]
    \ra{1.3}
    \centering
    \caption{The specification VReq.}\label{tabVAxisA}
    \begin{tabular}{lrrr}
      \toprule
        & VReq
        & VAxis
        & $S$\\
          \midrule
      States  & 12 & 15 & 15  \\
      Trans.  & 44 & 50 & 36  \\
      Events  &  9 & 11 & 11  \\
      \bottomrule
    \end{tabular}
  \end{table}

  Similarly, the specification HReq contains 19 events that are shared only with the plant HAxis. The events of HReq are the only observable events.
  The results for specification HReq are summarized in Table~\ref{tabVAxisB}.

  \begin{table}[ht]
    \ra{1.3}
    \centering
    \caption{The specification HReq.}\label{tabVAxisB}
    \begin{tabular}{lrrr}
      \toprule
        & HReq
        & HAxis
        & $S$\\
          \midrule
      States  & 112 & 128  & 80 \\
      Trans.  & 736 & 1002 & 320 \\
      Events  &  19 & 20   & 20 \\
      \bottomrule
    \end{tabular}
  \end{table}

  The specification HVReq contains ten events that are shared with the plants VAxis, HAxis, and HVN. Therefore, the set $H=\{\text{VAxis},\text{HAxis},\text{HVN}\}$. Since the plant $H$ is modular, we use the technique of Section~\ref{subsectionCMDES} and the algorithms of Komenda and Masopust~\cite{SCL12} to compute a set $\Sigma_{\kappa}$, with 14 events, that makes the specification, which is obtained by lifting HVReq to the set of events occurring in VAxis, HAxis, or HVN, conditionally decomposable. To apply Corollary~\ref{cor19}, we set the events of $\Sigma_{\kappa}$ and the events occurring in the specification HVReq observable, which results in the set of observable events $\Sigma_o$ with 16 events. We construct the coordinator $G_\kappa$ with 160 states, 1287 transitions, and 14 events, and three local nonblocking and maximally permissive controllable and normal supervisors $S_1$, $S_2$, $S_3$. The results are summarized in Table~\ref{tabHVAxis2A}.

  \begin{table}[h]
    \ra{1.3}
    \centering
    \caption{The specification HVReq.}\label{tabHVAxis2A}
    \begin{tabular}{lrrrrrrrr}
      \toprule
        & \multicolumn{1}{c}{$K$}
        & \multicolumn{3}{c}{$H$}
        & \multicolumn{3}{c}{} \\
        \cmidrule(lr){2-2}
        \cmidrule(lr){3-5}
        & HVReq
        & VAxis & HAxis & HVN
        & $S_1$ & $S_2$ & $S_3$\\
          \midrule
      States  & 7  & 15 & 128  & 1 & 516 & 1132 & 283 \\
      Trans.  & 35 & 50 & 1002 & 1 & 3395 & 10298 & 1692 \\
      Events  & 10 & 11 & 20   & 1 & 21 & 25 & 14 \\
      \bottomrule
    \end{tabular}
  \end{table}

  For comparison, the global plant $\text{VAxis}\, \|\, \text{HAxis}\, \|\, \text{HVN}$ has 1920 states, 23350 transitions, and 32 events, and the nonblocking and maximally permissive controllable and normal supervisor of the specification obtained by lifting HVReq to the alphabet of $\text{VAxis}\, \|\, \text{HAxis}\, \|\, \text{HVN}$ has 2064 states and 20120 transitions; see Section~\ref{sec:summary} for a summary.

  Finally, the specification UIReq contains 21 events that are shared with all four plants VAxis, HAxis, HVN, and UI. Again, the plant $H=\{\text{VAxis}, \text{HAxis}, \text{HVN}, \text{UI}\}$ is modular with the global specification UIReq, and hence we compute a set $\Sigma_{\kappa}$ consisting of 10 events, all of which occur in UIReq. Consequently, the set of observable events is formed by the alphabet of UIReq. We now compute the coordinator $G_{\kappa}$ with 4 states and 30 transitions, and four local nonblocking and maximally permissive controllable and normal supervisors $S_1$, $S_2$, $S_3$, $S_4$. The results are summarized in Table~\ref{tabUIReqA}.

  \begin{table}[h]
    \ra{1.3}
    \centering
    \caption{The specification UIReq.}\label{tabUIReqA}
    \begin{tabular}{@{}lrr@{~~~}r@{~~~}r@{~~~}r@{~~~}r@{~~~}r@{~~~}r@{~~~}r@{}}
      \toprule
        & \multicolumn{1}{c}{$K$}
        & \multicolumn{4}{c}{$H$}
        & \multicolumn{4}{c}{} \\
        \cmidrule(lr){2-2}
        \cmidrule(lr){3-6}
        & UIReq
        & VAxis & HAxis & HVN & UI
        & $S_1$ & $S_2$ & $S_3$ & $S_4$ \\
          \midrule
      States  & 256  & 15 & 128  & 1 & 2 & 432 & 768 & 12 & 96 \\
      Trans.  & 2336 & 50 & 1002 & 1 & 15 & 3488 & 6652 & 74 & 808\\
      Events  & 21 & 11 & 20 & 1 & 9 & 21 & 24 & 10 & 16 \\
      \bottomrule
    \end{tabular}
  \end{table}

  For comparison, $\text{VAxis}\, \|\, \text{HAxis}\, \|\, \text{HVN}\, \|\, \text{UI}$ has 3840 states, 75500 transitions, and 41 events, and the minimal automaton realizing a nonblocking and maximally permissive controllable and normal supervisor of the specification obtained by lifting UIReq to the alphabet of $\text{VAxis}\, \|\, \text{HAxis}\, \|\, \text{HVN}\, \|\, \text{UI}$ has 211200 states and 2751680 transitions.

  \subsection{Experimental Time Complexity}
  From the experimental time-complexity viewpoint, the required times in seconds to perform all computations, including the input/output operations and the minimization of the constructed automata, are summarized in Table~\ref{tabTime}.

  In addition, for specifications HVReq and UIReq, we further include, in parentheses, the time of the computation of the global supervisor constructed for the modular systems $H=\{\text{VAxis},\text{HAxis},\text{HVN}\}$ and $H=\{\text{VAxis},\text{HAxis},\text{HVN},\text{UI}\}$, respectively. In particular, the computation for UIReq and the corresponding modular plant $H=\{\text{VAxis},\text{HAxis},\text{HVN},\text{UI}\}$ allocated more than 10 GB of memory in ca. 10 minutes, and ran out of memory (oom) in hour and five minutes.

  \begin{table}[h]
    \ra{1.3}
    \centering
    \caption{Experimental time complexity in seconds.}\label{tabTime}
    \begin{tabular}{lrrrrrr}
      \toprule
        & VReq
        & HReq
        & HVReq (global)
        & UIReq (global)
        & Mono\\
      \midrule
      Time & 0.01 & 0.03 & 2.66 (9.37) & 2.72 (oom) & 4006 \\
      \bottomrule
    \end{tabular}
  \end{table}

  Finally, the computation of the nonblocking and maximally permissive controllable and normal supervisor constructed for the global specification $\text{VReq}\, \|\, \text{HReq}\, \|\, \text{HVReq}\, \|\, \text{UIReq}$ and the monolithic plant $\text{VAxis}\, \|\, \text{HAxis}\, \|\, \text{HVN}\, \|\, \text{UI}$ took more than one hour ($\approx 4006$ seconds). For completeness, we have computationally verified that the parallel composition of all the constructed local supervisors results in the nonblocking and maximally permissive monolithic supervisor.

  \subsection{Summary}\label{sec:summary}
  We have constructed nine local supervisors. The total time of the computations and the overall size of the constructed local supervisors are summarized in Table~\ref{tabSumB} (first column). All the considered local supervisors are nonblocking and maximally permissive in the sense that the resulting closed-loop system coincides with the nonblocking and maximally permissive monolithic closed-loop system.
  For comparison, we have included the monolithic approach (second column), and four monolithic approaches, one for each part of the specification (third column). The last column overviews the time and overall size of the high-level supervisors constructed by the hierarchical approach of Komenda and Masopust~\cite{firstPart}.

  \begin{table}[h]
    \ra{1.3}
    \centering
    \caption{The summary of results.}\label{tabSumB}
    \begin{tabular}{lrrrrr}
      \toprule
        & $9 \times \text{local}$
        & Monolithic
        & $4 \times \text{global}$
        & $4 \times \text{high}$ \\
      \midrule
      States &  3334 &  68672 &  213359 & 3768\\
      Trans. & 26763 & 616000 & 2772156 & 31486\\
      Time   & 5.42  & 4006   & oom     & ca. 11\\
      \bottomrule
    \end{tabular}
  \end{table}

\section{Conclusions}
  We investigated supervisory control of modular discrete-event systems under partial observations.
  We showed that the concept of hierarchical supervisory control called modified observation consistency (MOC) can be used to guarantee that the global nonblocking and maximally permissive normal supervisor can be achieved locally as the parallel composition of local normal supervisors.
  We considered the case of local specifications as well as the case of global specifications.

  We further showed that the global and local computations of nonblocking and maximally permissive normal supervisors coincide under the condition that all shared events are observable.
  This condition is stronger than MOC and nicely complements a similar condition of modular supervisory control under complete observation that all shared events are controllable.

  We illustrated our results on an industrial case study of the patient table of an MRI scanner.

  Finally, we would like to point out that it is worth combining both the modular approach and the hierarchical approach. This combination is particularly useful if the specification describes the required behavior in terms of high-level events, and it will very likely bring further improvements. However, this problem requires further investigation and we plan to discuss it in our future work.

\section*{Acknowledgment}
  This research was partially supported by the M\v{S}MT under the INTER-EXCELLENCE project LTAUSA19098, and by the Czech Academy of Sciences under RVO~67985840.

\bibliographystyle{IEEEtran}
\bibliography{biblio_journal}

\vfill

\end{document}